\documentclass[aps,pra,amsfonts,amsmath,amssymb,bm,twocolumn,multicol]{revtex4}
\begin{document}
\flushbottom

\title{Low-temperature specific heat of real crystals: 
Possibility of \\leading contribution of optical and short-wavelength acoustical vibrations}
\author{A. Cano and A.P. Levanyuk}
\affiliation{Departamento de F\' \i sica de la Materia Condensada C-III, Universidad Aut\' onoma de Madrid, E-28049 Madrid, Spain}
\date{\today}

\begin{abstract}
We point out that the repeatedly reported glass-like properties of crystalline materials are not necessarily associated with localized (or quasilocalized) excitations. In real crystals, optical and short-wavelength acoustical vibrations remain damped due to defects down to zero temperature. If such a damping is frequency-independent, e.g., due to planar defects or charged defects, these optical and short-wavelength acoustical vibrations yield a linear-in-$T$ contribution to the low-temperature specific heat of the crystal lattices. At low enough temperatures such a contribution will prevail over that of the long-wavelength acoustical vibrations (Debye contribution). The crossover between the linear and the Debye regime takes place at $T^* \propto \sqrt N$, where $N$ is the concentration of the defects responsible for the damping. Estimates show that this crossover could be observable. 
\end{abstract}

\maketitle

It is generally accepted that, at low enough temperatures, the specific heat of nonmetalic solids follows the Debye law: $C \propto T^3$ \cite{Landau_SP}. Loosely speaking, it can be said that the Debye law arises because of the thermal activation of the long-wavelength acoustical vibrations according to the Bose-Einstein distribution. Optical and short-wavelength acoustical vibrations do not contribute significantly to the low-temperature specific heat because their thermal activation is exponentially suppressed. Exception are glasses, for which a linear-in-$T$ dependence is known since long ago. Such a linear dependence is usually ascribed to the existence of the tunneling two-level systems postulated by Anderson {\it et al.} and, independently, Phillips in the early seventies \cite{Anderson72}. Up to date, however, the true microscopic nature of these two-level systems is unclear. A glass-like specific heat is observed in strongly disordered crystals, what upholds the common belief that in these crystals the above mentioned systems there also exist. ``Anomalous'' low-temperature specific heats were also observed in a number of crystals with relatively small concentrations of defects, but not in more perfect crystals of the same composition ---not in the same temperature range at least (see e.g. Refs. \cite{Lowless76,Ackerman81}, and the references therein). Appart from the evident conclusion that these anomalies are related to defects, no reliable explanation was found.

In this Letter, we argue that \emph{a linear-in-$T$ specific heat is a natural low-temperature property of any crystal which contains a small concentration of defects}. The key points are i) that the optical and short-wavelength acoustical vibrations of real crystals remain damped down to zero temperature due to presence of defects and ii) that the contribution to the specific heat of these damped vibrations may be linear-in-$T$ at low enough temperatures. In the view of this, it is evident that the contribution to the specific heat associated with these optical and short-wavelength acoustical vibrations may prevail over the (acoustic) Debye one at the lowest temperatures. The rest of the Letter is devoted to a more detailed argumentation and to estimations. 

Nowadays, the study of quantum effects in dissipative systems is a topic of vivid interest. As a result of this activity, the specific thermodynamic features of the damped harmonic oscillator are now well understood (see, e.g., Ref. \cite{Weiss}). It is worth mentioning that these features have already been proved relevant to explain some phenomena peculiar to vortex lattices in superconductors. We refer, in particular, to the significant contribution that vortons yield in the corresponding low-temperature specific heat \cite{Bulaevskii93}: vortons are acoustic phonons in what concerns to the dispersion law (they are connected to acoustic vibrations of the vortex lattice), but optic ones in what concerns to their finite damping at $\mathbf k =0$ (resulting from the vortex viscosity). It is worth mentioning that in incommensurate phases a similar situation takes place: phasons in incommensurate phases are analogous to vortons in superconductors. The low-temperature specific heat observed in some incommensurate phases has also a peculiar dependence on temperature \cite{Etrillard96}. Discussion on the origin of such a dependence will be published elsewhere \cite{Cano}. Here we shall concentrate on the low-temperature specific heat of ``ordinary'' crystal lattices. 

A detailed discussion on the free energy of a damped oscillator can be found e.g. in Ref. \cite{Weiss}. To our purposes, it is convenient to start by considering the case of a frequency-independent damping. The asymptotic low-temperature expansion of the free energy then reads
\begin{align}
F\approx F_0 - {\pi\over 6}{\gamma \over \omega_0}
{(k_BT)^2\over \hbar \omega_0}\left[
1+{2\pi^2\over 5}\left({k_BT\over \hbar \omega_0}\right)^2\right],
\label{F_oscil}\end{align}
where $F_0$ is the ground state energy, $\gamma$ is the damping coefficient, and $\omega_0$ is the natural frequency of the oscillator \cite{note_underdamped}. It is worth noticing the strong difference between this power-law asymptotic expansion and the exponential one that, in accordance with the Bose-Einstein distribution, is obtained for an undamped oscillator \cite{Landau_SP}. This is connected to the broadening of the density of states that the damping provokes (the broadening is such that there is no gap above the ground state, see Ref. \cite{Hanke95}). 

As we have mentioned, all vibrations of a real crystal (acoustical and optical ones) are in principle damped due to defects down to zero temperature. Let us assume for a while that this damping does not depend on the frequency as it is considered in Eq. \eqref{F_oscil}. At first glance, in accordance with this expression, it will be necessary to take into account absolutely all vibrational modes of the system when further computing, for instance, the corresponding low-temperature specific heat. But a basic assumption when obtaining Eq. \eqref{F_oscil} itself is that a certain separation between ``relevant'' and ``irrelevant'' degrees of freedom was possible: it is necessary that a part of the system acts as a reservoir to produce dissipation (damping) \cite{Weiss}. If Eq. \eqref{F_oscil} is used, some sort of ``double counting'' then seems unavoidable. However this is not completely true. Notice that, for a given finite temperature, there always exist acoustical vibrations with small enough frequencies for which the power-law expansion \eqref{F_oscil} makes no sense \cite{note_sense}. As a result, these acoustical vibrations with small wavevectors give rise, in particular, to the well known Debye contribution to the specific heat. They are just these acoustical vibrations which act as the above mentioned reservoir necessary to provide the damping (see below). 

The asymptotic expansion Eq. \eqref{F_oscil} can then be used to further estimate the contribution to the specific heat due to ``optical'' vibrations (optical and short-wavelength acoustical ones) in the low-temperature limit. Indeed at temperatures $T \ll \hbar \Delta/k_B$, where $\Delta $ is the characteristic frequency of the corresponding ``optical'' branch, it is sufficient to retain the lowest-order term in $T$. Thus, \emph{the contribution to the low-temperature specific heat of a damped ``optical'' branch} can be roughly estimated as
\begin{align}
C_\text{op}
\sim {k_B\over d^3}{\gamma\over \Delta }{k_BT\over \hbar \Delta},
\label{C_whole_branch}\end{align}
where $d$ is the characteristic interatomic distance. Here the damping constant $\gamma$ has been assumed to be the same for the whole ``optical'' branch. This ``optical'' contribution to the specific heat will prevail over the acoustic one \cite{Landau_SP} at $T\lesssim T^*\simeq (\gamma/\Delta)^{1/2} \Theta$,
where $\Theta = \hbar c^{3/2}/(k_B \Delta^{1/2} d^{3/2})$, with $c$ being the velocity of sound, is of the order of magnitude of the Debye temperature. 

Let us mention that it is not completely consistent to single out one ``optical'' branch and the acoustic reservoir given that there is a ``viscous'' coupling between these ``optical'' vibrations due to the damping (described by non-diagonal terms in the corresponding dissipative function). However, in order to reveal the linear-in-$T$ contribution to the low-temperature specific heat and to estimate this contribution in order of magnitude, this neglection seems permissible (in the limit of small defect concentration the cross terms refer to a finite number of ``optical'' vibrations: with the same $\mathbf k$ but belonging to different branches). This imprecision is the price we have to pay for the treatment of a complicated but hermitian many-body problem in terms of a set of one-degree-of-freedom but non-hermitian subsystems. Anyway, a valuable information has been obtained from this latter approach: after solving the mentioned many-body problem, the corresponding density of states one must find should be similar to that of a damped oscillator \cite{Hanke95}. 

For further progress, it is necessary to estimate the corresponding damping. This estimation, however, is far from being trivial even from experimental data. Inelastic scattering and nuclear magnetic resonance experiments, for instance, permit some estimations. But one must bear in mind that these experiments provide a partial information. From the former it can only be extracted the high-frequency damping. From the latter it is the low-frequency one, but ``averaged'' over wavevectors. The damping of polar optic modes, in particular, can also be estimated from experimental data on dielectric losses. But these estimates will not be complete: dielectric losses are connected to the damping of the long-wavelength optic modes only. The theory on these dielectric losses is well documented, so let us take advantage of this point to illustrate that \emph{the defect-induced damping does not vanish at zero temperature} and to make some estimates. 

The problem of the dielectric losses in ferroelectrics due to symmetry-breaking defects was theoretically studied e.g. in Ref. \cite{Balagurov72} and reviewed in Ref. \cite{Gurevich91}. The mechanism of loss considered was the defect-permitted radiation of acoustic waves when applying a time-dependent homogeneous electric field (a symmetry-breaking defect, e.g. an interstitial atom, induces a local linear coupling between the electric field, or polarization, and the strain). As a result one finds that the low-frequency damping constant of a polar mode can be inferred as 
\begin{align}
\gamma (\omega \to 0) \sim Nd^3 \omega_D^{1-n}\omega^{n}.
\label{gamma_omega_SBdefect}\end{align}
Here $N$ is the generalized concentration of defects \cite{nota_defect_conc}, $\omega_D$ is the Debye frequency and $n = 0,1,2$ for planar, linear and point defects respectively. (The defects are considered strong defects in the same sense that in Ref. \cite{Gurevich91}.) The same results are obtained for non symmetry-breaking defects \cite{Balagurov72,Silverman62}.

It is worth mentioning that the basic ingredient to obtain this defect-induced losses is the accounting for the local defect-induced changes in the properties of the corresponding crystal, i.e., the local symmetry breaking and/or the local changes in the material constants (inhomogeneities). This is sufficient to ``connect'' the given vibrational mode with the acoustic reservoir, which further leads to the corresponding damping (already within the approximation of small defect concentration). Hence this defect-induced damping is not restricted to polar modes: in principle any vibration is affected by these defect-induced local changes; and, consequently, Eq. \eqref{gamma_omega_SBdefect} is expected to be valid for non polar modes also. Let us stress that the role played by defects, via these local changes, is simply to permit the above mentioned connection between the oscillators and the reservoir (no additional degrees of freedom associated with any ``internal defect dynamics'' are considered). 

As we see in Eq. \eqref{gamma_omega_SBdefect}, a frequency-independent damping of the ``optical'' vibrations is obtained due to planar defects ($n=0$). According to the exposed above, this further gives a linear-in-$T$ contribution to the specific heat which is predominant at low enough temperatures. The temperature crossover between this linear-in-$T$ regime and the Debye one is $T^*\sim (Nd^3)^{1/2}\Theta=(N_\text{planar}d)^{1/2}\Theta$. For a concentration of planar defects corresponding to the typical dimensions of the crystal blocks, i.e. one interblock boundary per $\rm \mu m$ approximately, this crossover temperature $T^*$ results to be of a few Kelvins. 

The presence of charged defects in ionic crystals provide another mechanism of damping for the corresponding polar modes. The frequency dependence of the dielectric losses expected in this case was analyzed in Ref. \cite{Vinogradov61,Schlomann64}. This analysis can be easily generalized in order to compute the frequency dependence of the damping of a polar mode with ${\mathbf k} \not = 0$ in the case of a realistic charge density distribution. (The choice of the charge density distribution in Ref. \cite{Vinogradov61,Schlomann64} was somewhat arbitrary, see Appendix for further details). As a result, the damping coefficient of practically all the polar modes turns out to be frequency-independent in the limit $\omega \to 0$. This leads to a linear-in-$T$ specific heat at low temperatures, the orders of magnitude similar to the obtained above. 

In the case of a frequency-dependent damping such that $\gamma \propto \omega$, e.g. for linear defects [$n=1$ in Eq. \eqref{gamma_omega_SBdefect}, see Refs. \cite{Balagurov72,Silverman62}], the damping merely reduces to a renormalization of the optic masses. So in this case, the contribution of ``optical'' vibrations will be the ``ordinary'' one.  

Let us finally mention that $\gamma \propto \omega ^2$ can occur due to both, point defects [$n=2$ in Eq. \eqref{gamma_omega_SBdefect}, see Refs. \cite{Balagurov72,Silverman62}] and strongly-correlated charged defects in the case of polar modes (see Refs. \cite{Vinogradov61,Schlomann64}). Following e.g. Ref. \cite{Weiss} it can be seen that this further gives a Debye-like contribution ($\propto T^3$) to the low-temperature specific heat \cite{note_page}. Let us stress that this Debye-like contribution has a completely different origin than the usual (acoustic) Debye one: it comes from damped ``optical'' vibrations. It is small because of the smallness of the defect concentration we are considering. But one can speculate that, for high enough concentration of defects, this contribution may be comparable to that of the acoustic phonons. 

As an experimental example in which the low-temperature contribution to the specific heat due to damped ``optical'' vibrations has probably been observed, it is worth mentioning the low-temperature anomaly of Li$_3$N reported in Ref. \cite{Ackerman81}. This anomaly cannot be interpreted as a proper glass one (on the basis of the tunneling two-level model, for instance) because such a interpretation would be inconsistent with the thermal conductivity data: the magnitude of this conductivity is too small and its temperature dependence is somewhat different from that expected in a proper glass (no plateau is observed at $T\gtrsim 10\, \rm K$) \cite{Ackerman81}. But, unfortunately, we are not aware of the complementary experiments mentioned above necessary to confirm (or to reject) that the low-temperature specific heat observed in this particular case is due to the damping of the corresponding ``optical'' vibrations. Mention that these experiments (inelastic scattering, nuclear magnetic resonance, etc.) must be realized for the same sample because the anomaly is expected to be related to the presence of defects. Anyway let us notice that this low-temperature specific heat anomaly, such that $C\sim 2\times 10^{-5}\rm \, J\, cm^{-3}\, K^{-1}$ at $T\simeq 1\, \rm K$ \cite{Ackerman81}, can be reproduced, in order of magnitude, from Eqs. \eqref{C_whole_branch} and \eqref{gamma_omega_SBdefect} with $n=0$, a concentration of defects $Nd^3\sim 10^{-4}$ (which could correspond to one interblock boundary per $\rm \mu m$, i.e. $N=N_\text{planar}d^{-2}\sim 10^{-4}d^{-3}$), and the typical values $\hbar\omega_D/k_B,\hbar\Delta/k_B \sim 100\, \rm K$.

In conclusion, we have shown that \emph{the low-temperature specific heat of the real crystals may have a significant contribution due to damped optical and short-wavelength acoustical vibrations}. Such a contribution may split into linear-in-$T$ and Debye-like terms. The former will prevail over the Debye contribution of acoustical vibrations for temperatures $\lesssim 1\,\rm K$ for typical concentrations of defects of nominally pure crystals. The latter might be comparable with the Debye one for high enough concentration of defects. Let us stress that these contribution are exclusively due to damped excitations. This damping is due to defects, but defects themselves do not introduce any additional degrees of freedom in our considerations. If defect excitations are taken into account, some additional contribution to the specific heat will be obtained similar to that reported for glasses. Mention also that, as we have shown, the low-temperature properties of a real system may arise from nearly all of its phononic modes (all ``optical'' vibrations of a real crystal contribute to the corresponding low-temperature specific heat as we have seen). This does not contradict the general point of view according to which, even in glasses, it is accepted that only a small number of phononic modes, i.e. the long-wavelength acoustic ones, are relevant to the corresponding low-temperature properties. But it indicates that the low-energy excitations responsible for non-Debye contributions already exist in nominally perfect crystals, having their origin in the phononic normal modes.

We wish to thank S. Vieira, F. Guinea and especially M.A. Ramos for useful discussions.

\appendix

\section{D\lowercase{amping of a polar mode due to a disordered charge distribution}}

Let us calculate the $\omega$ dependence of the damping constant $\gamma$ of a polar mode due to the presence of defects that induce a disordered charge distribution in the medium. This can be obtained, in the low-frequency regime we are interested in, from the $\omega$ dependence of the imaginary part of the susceptibility $\hat \chi (\omega,{\mathbf k})$: $\gamma(\omega\to 0 , {\mathbf k})\sim \omega^{-1}\hat \chi''(\omega\to 0 , {\mathbf k})$. In the following we shall consider that the charge density distribution is correlated in accordance with the Debye-H$\rm \ddot u$ckel theory (see, e.g., Ref. \cite{Martin88}). The results obtained in such a way are different from that given in Refs. \cite{Vinogradov61,Schlomann64
}, where the choice of the charge density distribution was somewhat arbitrary.

\onecolumngrid

The $\mathbf k$-th Fourier component of the polarization can be written as
\begin{align}
P_i (\omega, {\mathbf k}) = 
\chi_0(\omega,{\mathbf k})E_i(\omega,{\mathbf k}) 
+ {1\over V }\iint \langle \rho({\mathbf r})u_i ({\mathbf r},t)\rangle
e^{-i{\mathbf k}\cdot {\mathbf r}}e^{i{\omega t}}d{\mathbf r} dt,
\label{PolaP}\end{align}
where $\chi_0$ is the (scalar) susceptibility in absence of defects (the medium is assumed to be isotropic), $V$ is the volume of the system, $\rho$ is the disordered charge distribution due to the presence of the defects, $\mathbf u$ is the acoustic displacement vector, ${\mathbf E}$ is the electric field, and $\langle \dots \rangle$ denotes statistical average. The $i$-th component of this displacement vector can be written as \footnote{In real space, the equation of motion for the displacement vector can be written as $\ddot {\mathbf u} = c_t^2 \nabla^2 {\mathbf u}+ (c_l^2- c_t^2)\nabla (\nabla \cdot {\mathbf u}) + \mu^{-1} \rho {\mathbf E}$, see Ref. \cite{Schlomann64}.} 
\begin{align}
u_i(\omega,{\mathbf k})=-\mu^{-1}\Lambda_{ij}^{-1}(\omega,{\mathbf k})
\sum_{\mathbf k'}\rho({\mathbf k}-{\mathbf k'})E_j(\omega,{\mathbf k'}), 
\end{align}
where $\mu $ is the density of the medium and 
\begin{align}
\Lambda_{ij}^{-1}(\omega,{\mathbf k})
={1\over (\omega^2 - c_l^2 k^2)(\omega^2 - c_t^2 k^2)}\big[(\omega^2 - c_l^2 k^2)\delta_{ij}+(c_l^2 - c_t^2 )k_ik_j\big],
\end{align}
with $c_l$ and $c_t$ being the longitudinal and transversal velocities of sound respectively. Substituting this expression in Eq. \eqref{PolaP} we get 
\begin{align}
P_i (\omega, {\mathbf k}) = &
\chi_0(\omega,{\mathbf k})E_j(\omega,{\mathbf k}) 
- \mu^{-1} \negthickspace\sum _{{\mathbf k'}}
K({\mathbf k}-{\mathbf k'})
\Lambda_{ij}^{-1}(\omega,{\mathbf k'})
E_j(\omega,{\mathbf k}),
\label{P_E}\end{align}
where $K({\mathbf k})$ is defined as the spectral density of the charge distribution:
\begin{align}
\langle \rho({\mathbf k})\rho({\mathbf k'}) \rangle \propto K({\mathbf k}) 
\delta_{{\mathbf k},-{\mathbf k'}}.
\label{correlation}\end{align}

In accordance with Eq. \eqref{P_E}, the susceptibility due to the defects can be written as
\begin{align}
\chi_{ij,\text{def}} (\omega,{\mathbf k})= -\mu^{-1} \negthickspace\sum _{{\mathbf k'}}K({\mathbf k}-{\mathbf k'})\Lambda_{ij}^{-1}(\omega,{\mathbf k'}). 
\label{}\end{align}
After integration over wavevectors the non-diagonal terms in $\chi_{ij,\text{def}}$ vanish (we are considering an isotropic medium). So we actually have $\chi_{ij,\text{def}}= \chi_\text{def} \delta_{ij}$, where
\begin{align}
\chi_\text{def} (\omega,{\mathbf k})= -\mu^{-1} \negthickspace\sum _{{\mathbf k'}}
K({\mathbf k}-{\mathbf k'})\Lambda_{ii}^{-1}(\omega,{\mathbf k'}) 
\label{chiii}\end{align}
(here double subscript does not imply summation).

In accordance with the Debye-H$\rm \ddot u$ckel theory (see, e.g., Ref. \cite{Martin88}), the charge density around the point ${\mathbf r}=0$ at which the charge density is known, say $\rho (0)$, is such that 
\begin{align}
\langle \rho({\mathbf r \not = 0}) \rangle _{\rho (0)}\sim r^{-1}\exp(-r/r_D),
\end{align}
where $r_D$ is the Debye screening length: $r_D = [k_BT/(8\pi e^2 N)]^{1/2}$, where $N$ is the defect concentration and $T$ can be taken as the annealing temperature in our case (the charge distribution is assumed to be in thermal equilibrium at this temperature). The desired correlation function $\langle \rho({0})\rho({\mathbf r}) \rangle$ can be obtained by averaging the above magnitude over all the possible $\rho(0)$. This gives 
\begin{align}
\langle \rho({0})\rho({\mathbf r}) \rangle 
= {N e^2}\Big[ \delta(\mathbf r) - {1\over 4\pi r_D^2 r}\exp(-r/r_D)\Big].
\end{align}
The spectral density of the defect charge distribution is therefore $K({\mathbf k}) = {\mathcal K}{r_D^2k^2 /(1+r_D^2k^2)}$,
where $\mathcal K = Ne^2/V$. 

The main contribution to the integral in Eq. \eqref{chiii} then comes from the poles of $\Lambda_{ii}^{-1}(\omega,{\mathbf k'})$. We are interested in the case of small frequencies. Taking into account that the poles of $\Lambda_{ii}^{-1}(\omega,{\mathbf k'})$ are such that $k'\sim \omega/c$ ($c\sim c_l,c_t$), we further can distinguish the following cases. For small wavevectors ($k\ll \omega/c$), in the integrand of Eq. \eqref{chiii} we can take
\begin{align}
K({\mathbf k} - {\mathbf k}')\approx 
{\mathcal K}{r_D^2k'^2 \over 1+r_D^2k'^2}\sim
\begin{cases}
{\mathcal K}r_D^2k'^2,& r_D k' \ll 1,\\
{{\mathcal K}},& r_D k' \gg 1.\\
\end{cases}\end{align}
We then have 
\begin{align}
\chi_\text{def}'' (\omega \to 0,k\ll \omega/c)\propto
\begin{cases}
Nd^3(r_D/d)^2(\omega/\omega_D)^3, & \omega \ll c r_D^{-1},\\
Nd^3(\omega/\omega_D), & \omega \gg c r_D^{-1},\\
\end{cases}\end{align}
where $\omega_D\sim c d^{-1}$ is the Debye frequency. The $\omega$ dependence in these expressions coincides with that reported in Refs. \cite{Vinogradov61,Schlomann64
}. For large wavevectors ($k\gg \omega/c$), in the integrand of Eq. \eqref{chiii} we can take
\begin{align}
K({\mathbf k} - {\mathbf k}')\approx 
{\mathcal K}{r_D^2k^2 \over 1+r_D^2k^2}.
\end{align}
We then have 
\begin{align}
\chi_\text{def}'' (\omega \to 0,k\gg \omega/c)\propto
Nd^3(\omega/\omega_D).
\end{align}
From these expressions it follows that, as a result of the presence of charged defects, the damping coefficient of practically all the polar modes does not depend on the frequency in the limit $\omega \to 0$.

\twocolumngrid


\begin{thebibliography}{0}

\bibitem{Landau_SP} L.D. Landau and E.M. Lifshitz, {\it Statistical Physics} (Pergamon, Oxford, 1980).

\bibitem{Anderson72} P.W. Anderson, B.I. Halperin and C.M. Varma, Phil. Mag. {\bf 25}, 1 (1972); W.A. Phillips, J. Low Temp. Phys. {\bf 7}, 351 (1972). 

\bibitem{Lowless76} W.N. Lawless, Phys. Rev. Lett. {\bf 36} 478 (1976); Phys. Rev. B {\bf 14} 134 (1976); J.J. De Yoreo, R.O. Pohl and G. Burns, {\it ibid.} {\bf 32} 5780 (1985), R. Villar, E. Gmelin and H. Grimm, Ferroelectrics {\bf 69}, 165 (1986).

\bibitem{Ackerman81} D.A. Ackerman {\it et al.}, Phys. Rev. B {\bf 23} 3886 (1981).  

\bibitem{Weiss} U. Weiss, {\it Quantum Dissipative Systems} (World Scientific, Singapore, 1999).

\bibitem{Bulaevskii93} L.N. Bulaevskii and M.P. Maley, Phys. Rev. Lett. 71, 3541 (1993); G. Blatter and B.I. Ivlev, Phys. Rev. B 50, 10272 (1994); A.L. Fetter and S.R. Patel, {\it ibid.} 54, 16116 (1996); R. Iengo and C.A. Scrucca, {\it ibid.} 57, 6046 (1998).

\bibitem{Etrillard96} J. Etrillard et al., Phys. Rev. Lett. {\bf 76}, 2334 (1996); J. Etrillard et al., Europhys. Lett. {\bf 38}, 347 (1997).

\bibitem{Cano} A. Cano and A.P. Levanyuk, cond-mat/0404437 (unpublished).

\bibitem{note_underdamped} In Eq. \eqref{F_oscil} it is assumed that the oscillator is underdamped ($\omega_0 \gg \gamma$). If the oscillator is overdamped the term $\propto T^2$ does not change (notice that it does not depend on the mass of the oscillator), but $\omega_0^2$ must be replaced by $-3\omega_0^4/\gamma^2$ in the second term in square brackets if $\omega_0 \ll \gamma$.

\bibitem{Hanke95} A. Hanke and W. Zwerger, Phys. Rev. E {\bf 52}, 6875 (1995).

\bibitem{note_sense} For small enough wavevectors $\mathbf k$ acoustical vibrations are underdamped: $\gamma _\text{ac}/\omega_\text{ac}\propto k$ [see e.g. L.D. Landau and E.M. Lifshitz, {\it Fluid Mechanics}, (Pergamon, Oxford, 1987)]; so in principle Eq. \eqref{F_oscil} would apply \cite{note_underdamped}. However one has to realize that, for instance, the second term in square brackets becomes comparable to the first one for $\omega_0\lesssim k_BT/\hbar$. Consequently, for acoustical vibrations with small enough frequencies, the power-law expansion \eqref{F_oscil} makes no sense: higher-order terms cannot be neglected. The expression one finally obtains for these low-frequency acoustical vibrations is what follows from the Bose-Einstein distribution, i.e. their damping is unimportant.

\bibitem{Balagurov72} B.Ya Balagurov and R.O. Zaitsev, Fiz. Tverd. Tela {\bf 14}, 52 (1972). 

\bibitem{Gurevich91} V.L. Gurevich and A.K. Tagantsev, Adv. Phys. {\bf 40}, 719 (1991).

\bibitem{nota_defect_conc} The generalized defect concentration $N$ is just $N$ for point defects, $N=N_\text{linear}d^{-1}$ for linear defects with concentration $N_\text{linear} \, \rm [cm^{-2}]$, and $N=N_\text{planar} d^{-2}$ for planar defects with concentration $N_\text{planar} \,\rm [cm^{-1}]$.

\bibitem{Silverman62} B.D. Silverman, Phys. Rev. {\bf 125}, 1921 (1962).

\bibitem{Vinogradov61} V.S. Vinogradov, Fiz. Tverd. Tela {\bf 2}, 2622 (1961) [Sov. Phys.-- Solid State {\bf 2}, 2338 (1961)]; 

\bibitem{Schlomann64} E. Schl$\rm \ddot o$mann, Phys. Rev. {\bf 135}, A413 (1964); B.M. Garin, Fiz. Tverd. Tela {\bf 32}, 3314 (1990).
\bibitem{note_page} See Ref. \cite{Weiss}, p. 133.

\bibitem{Martin88} Ph. A. Martin, Rev. Mod. Phys. {\bf 60}, 1075 (1988).


\end{thebibliography}
\end{document}